\documentclass[reprint]{revtex4-1}
\usepackage{amsfonts,amssymb,amsmath,amsthm,graphicx}
\usepackage{url}
\usepackage{bbm}
\usepackage[usenames,dvipsnames]{xcolor}
\usepackage{xcolor}
\usepackage{fancyhdr}
\usepackage{fancyvrb}
\usepackage{bm}
\usepackage{setspace}
\usepackage{subfigure}
\usepackage{tabularx}
\usepackage{float}
\usepackage[normalem]{ulem}
\usepackage{comment}
\usepackage[colorlinks=true, linkcolor=black,  citecolor=black, urlcolor=black,linktocpage=true]{hyperref}

\begin{document}
\title{Long term test of a fast and compact Quantum Random Number Generator}
\author{D. G. Marangon}
\email{davide.marangon@crl.toshiba.co.uk}
\thanks{\\Accepted in IEEE/OSA Journal of Lightwave Technology \\\url{http://dx.doi.org/10.1109/JLT.2018.2841773}}
\author{A. Plews}
\author{M. Lucamarini}
\author{J. F. Dynes}
\author{A.W. Sharpe}
\author{Z. L. Yuan}
\author{A. J. Shields}
\affiliation{Toshiba Research Europe Ltd, Cambridge Research Laboratory, 208 Cambridge Science Park, Milton Road, Cambridge, CB4 0GZ, UK}

\begin{abstract}
Random numbers are an essential resource to many applications, including cryptography and Monte Carlo simulations. Quantum random number generators (QRNGs) represent the ultimate source of randomness, as the numbers are obtained by sampling a physical quantum process that is intrinsically probabilistic. However, they are yet to be widely employed to replace deterministic pseudo random number generators (PRNG) for practical applications. QRNGs are regarded as interesting devices. However they are slower than PRNGs for simulations and are typically seen as clumsy laboratory prototypes, prone to failures and unreliable for cryptographic applications. Here we overcome these limitations and demonstrate a compact and self-contained QRNG capable of generating random numbers at a pace of 8 Gbit/s uninterruptedly for 71 days. During this period, the physical parameters of the quantum process were monitored in real time by self-checking functions implemented in the generator itself. At the same time, the output random numbers were analyzed with the most stringent suites of statistical tests. The analysis shows that the QRNG under test sustained the continuous operation without physical instabilities or hardware failures. At the same time, the output random numbers were analyzed with the most stringent suites of statistical tests, which were passed during the whole operation time. This extensive trial demonstrates the reliability of a robustly designed QRNG and paves the way to its use in practical applications based on randomness.
\end{abstract}
\maketitle

\section{Introduction}
Random numbers are essential in many fields of science and information technology. The two main ways to generate them are either by iterating deterministic algorithm, the so-called pseudo random number generators (PRNG), or by sampling a natural physical process, the so-called true random number generators (TRNG). Although natural randomness has been acknowledged as the ideal method to generate random numbers for simulations or cryptographic keys \cite{1,2}, PRNGs are normally preferred for these tasks \cite{3,4}. Simulation-wise, the typical motivation is that TRNGs do not allow for results reproducibility. Cryptography-wise, TRNGs are believed to silently drift or break over time, thus compromising security.

On the other hand, PRNGs are intrinsically predictable and this can be exploited to break cryptographic security \cite{5,6,7}. Moreover, in Monte Carlo methods, the presence of artefacts or patterns in the strings generated by a PRNG can produce unreliable simulation results \cite{8}. Finally, reproducibility can be easily achieved with a TRNG by recording the output number stream and reusing it. It is reasonable to expect that such an approach will become increasingly viable as TRNGs' generation rates increases \cite{9,10}.

Motivated by this premise, we present in this work a TRNG that exploits the intrinsic randomness of a quantum optical process \cite{11,12} to meet both the reliability and the high generation rate demanded by applications. We demonstrate the stable and continuous generation of random numbers from this quantum random number generator (QRNG) at 8 Gbit/s rate. The outputted strings pass an extensive application of statistical randomness tests after a weak post-processing.

With the exception of few recent works, e.g. \cite{13,14,15}, ultrafast generation rates from a QRNG are typically achieved during proof of principle demonstrations. Lab equipment is used to generate the random signals, which are acquired by fast digital oscilloscopes \cite{16,17,18,19}. Then, analysis and post-processing of the recorded data are done ``offline", but the reported generation rates are evaluated as if all these processes were performed ``on the fly". In addition, the statistical tests to assess the quality of the numbers are applied just few times, typically once, on a limited data sample, mostly because of the oscilloscope finite storing capacity. Drawing a conclusion from this small statistics requires the additional strong assumption that the analyzed sample is representative of the entire stream the generator would normally output.

In this work, we present a self-contained small QRNG that performs live analysis and live post-processing of the random signals. The device is capable of stable operation and is self-checking. The latter feature means that QRNG monitors some parameters of its main internal components, to check whether they are operating properly. As a proof, we present the results of an uninterrupted test of a QRNG lasted 71 days. The QRNG's ultrafast generation rate allowed us to rapidly collect sufficient data to repeatedly apply the most stringent batteries of statistical tests several times.

In the following, we will review the QRNG's main features and present the results of the long term test.
\section{The generator}
The QRNG described in this work is based on the same working principle as the one presented in \cite{21}. However, rather than performing a proof-of-principle experiment, we integrated all the optical and electronic components into a stand-alone self-contained device, depicted in Figure 1.
\begin{figure}[t]
\includegraphics[width=\columnwidth]{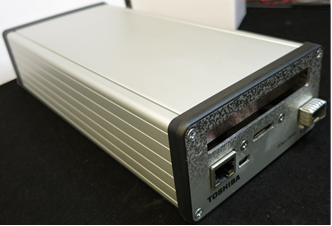}
 \caption{The QRNG self-contained unit. The compact enclosure's size is $10 \times 23 \times 5 \text{cm}^3$.}
 \label{fig1}
\end{figure}

The physical mechanism exploited to generate random numbers is the spontaneous emission from a pulsed laser \cite{21}. The physical core of the generator comprises a 1550 nm laser that emits steady state pulses with a repetition rate of 1 GHz. The use of this wavelength was motivated by the wide availability of high bandwidth lasers and photodiodes at telecom wavelengths. Therefore, the generator realization is eased since it can be built by employing standard commercial optical components. The pulses are emitted by a distributed feedback (DFB) laser diode in a standard 14-pin butterfly package integrating a photodiode for power monitoring, and a thermistor together with a thermoelectric cooling (TEC) for temperature regulation.

The delay of 1 ns between two consecutive pulses is sufficient to empty the laser cavity between two successive emissions. Therefore, each newly stimulated emission is triggered by a spontaneously emitted photon that carries the phase of the vacuum field, which is completely random over the interval $[0,2\pi]$.

This phase is then measured by interfering pairs of optical pulses in a one-bit-delayed fibre-based interferometer, whose optical output is sent to photodiode (PD). The PD is a commercial InGaAs/InP PIN receiver for classical optical communication, featuring a bandwidth of 5 GHz. The PD converts the random intensity optical input into a randomly varying current that is sampled by an analog-to-digital converter (ADC) with 10-bit resolution. These samples are therefore unpredictable by virtue of the physical uncertainty associated with the vacuum state.

This quantum uncertainty prevents a malevolent adversary to predict when a given number will be emitted. However, the adversary can still take advantage of the arcsine probability distribution of the interferometer output, and bet on the appearance of the most likely intensity values, which are the ones at the extremes of the distribution. It is therefore necessary to manipulate the numbers with a post-processing technique to remove this impairment. We use a finite impulse response (FIR) filter to unbias the numbers and to achieve a flat distribution of the 8-bit integers \cite{21,22}. This technique consists in transforming a raw integer sample $x(n)$ into an unbiased one $y(n)$, by means of the relation $y(n) = b_0 x(n)+b_1 x(n-1)+\dots + b_M x(n-M) \mod 2^8$, with binomial coefficients $b_i=M!/(i!(M-i)!)$. After the FIR processing, the QRNG outputs a stable stream of unbiased numbers at 1 GB/s.

It is worth clarifying that the FIR filter is not equivalent to a randomness extractor, which distills the entropy of the output string by compressing it \cite{23}. The filter does not perform compressing, but simply scrambles the raw samples by dispersing each input $x$ into $M+1$ outputs $y$ as a pragmatic unbiasing algorithm. Hence, the filtered output has exactly same entropy per bit as its raw input. We stress that the actual entropy in the FIR filtered output needs to be considered when used in sensitive cryptographic applications.

On the other hand, it is worth emphasizing that for applications where security is not a concern and generation speed is required, e.g. Monte Carlo methods, the FIR filter remains extremely effective. This holds when the initial amount of entropy in the generated strings is high, like in our case with average raw entropy of 0.9 bits. This is shown in Sec. V, where a large number of strict statistical tests are passed without applying temporal restrictions on the generated strings.

The compactness of the QRNG is achieved by embedding optical and electronic components into a printed circuit board (PCB). The functions of driving and monitoring the devices and of processing and outputting the signals are performed by an FPGA, which constitutes the electronic and digital backbone of the generator.

\begin{figure}[t]
\includegraphics[width=\columnwidth]{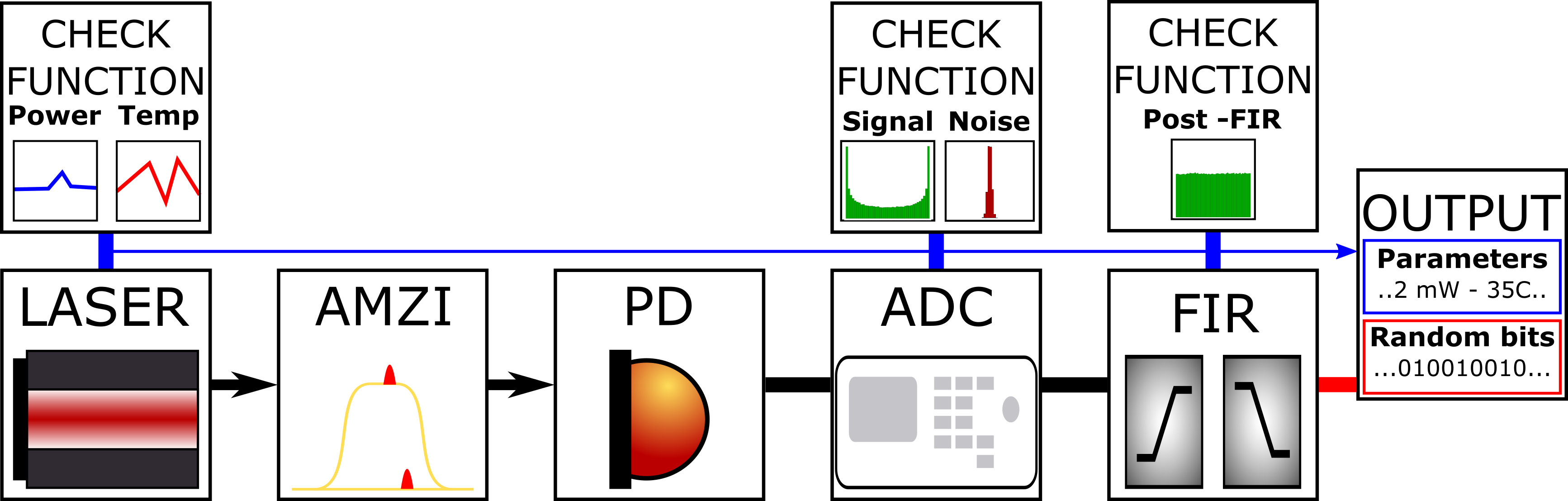}
\includegraphics[width=\columnwidth]{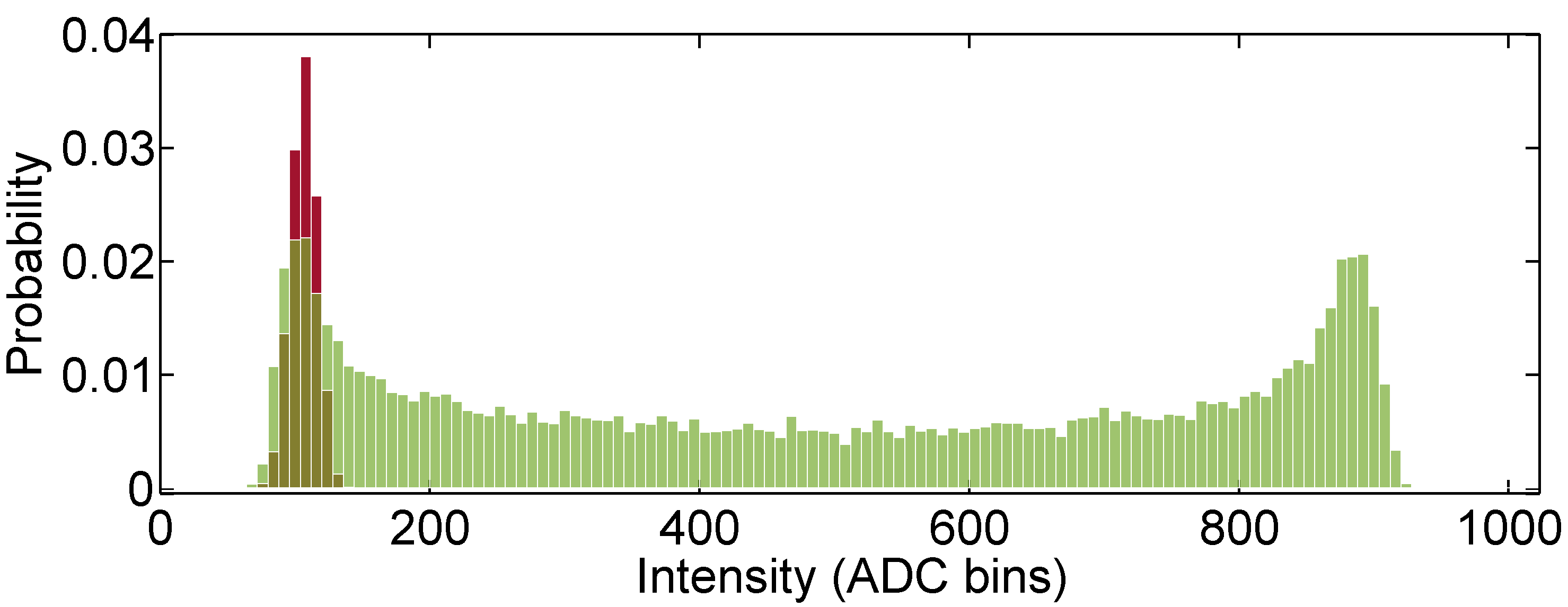}
 \caption{Top: Logic blocks of the QRNG. The bottom row represents the part of the device directly involved in the generation of the random numbers. The top row represents the digital functions that were implemented to monitor the system operation. Bottom: The foreground histogram (green) represents the interference intensity values measured by the PD and digitized by the ADC. The data are distributed according the expected arcsine distribution. The background histogram (red) represents the noise signal acquired between two pulses. Both these histogram are continuously acquired by the check functions implemented in the generator board.}
 \label{fig2}
\end{figure}

The versatility of the FPGA allows us to program different functions for the live monitoring of the physical parameters. This sort of sanity check is essential to discover instabilities or malfunctions that might affect the output randomness. In Figure 2, a logic scheme of the QRNG is reported with the main check functions reported.

The main physical active component is the laser. The laser output power determines the maximal range of the interference signal and hence the range of the current signal sampled by the ADC. This has to be finely tuned such that the interference signal can span all the available dynamical range without exceeding the upper digitization limit. To control power fluctuations due to the variation of laser temperature, we provided the laser with a temperature controller and implemented a function that monitors both the laser output power and the laser temperature. This way, the user can be aware of the laser status and can abort the generation in case of a sudden drift from the optimal operating condition.

The QRNG samples the PD output signal with a rate of 2 GS/s so that every 1 ns two data samples are generated. The first sample collects the optical interfering signal, the so-called foreground data, as depicted in the foreground (green) histogram in Figure 2 bottom. The second sample collects the non-optical signal between two interference pulses, i.e., the noise level of the PD (background data), as depicted in the background (red) histogram in Figure 2 bottom. The QRNG is programmed to continuously collect and histogram foreground and background data. This function is important for two reasons. On one hand, it allows the user to check whether the foreground distribution follows the theoretical intensity distribution pertaining to the interference process. On the other hand, it enables the monitoring of the background data, which is useful to detect a malfunction of the laser driving system. In particular, this makes it possible to detect light pulses emitted during the laser's off time, which could hinder the phase randomization of consecutive pulses.

In parallel to the histogram of the above physical distributions, the QRNG computes also the distribution of the data after the FIR, to enable a comparison against the uniform distribution expected from unbiased numbers.

A standard USB communication protocol lets the user set the operational parameters of the device and read the data produced by the monitor functions. High speed data interfaces, such as 10G and SATA, are also provisioned for outputting 8 Gb/s stream of random numbers.  In this study, the SATA interface is used to feed the random bit stream to a PC for statistical randomness tests.

\begin{table}
\begin{tabularx}{\columnwidth}{p{1 cm} p{2.5 cm} p{2.5 cm} p{2.5 cm}}
&\textbf{Parameter} & \textbf{Value} & \textbf{Unit}\\\hline\hline
&Laser Power & 5.03 $\pm$ 0.01 & mW\\\hline
\end{tabularx}
\caption{Laser output power (mW) monitored by the health check functions of the QRNG.}
\label{tab1}
\end{table}

\begin{figure}[t]
\includegraphics[width=\columnwidth]{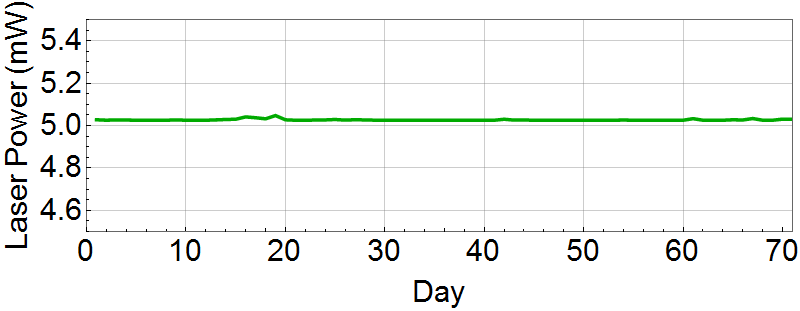}
 \caption{Daily average values of the laser powers logged by the power monitoring function.}
 \label{fig3}
\end{figure}
\section{Testing strategy}
To assess the QRNG suitability to cryptographic applications, we studied its response over a long period of time. In particular we focused on the two following properties:

1)	\textsl{Operational stability}:  the operational parameters of the components, such as the output power of the laser or its temperature, are required to not deviate from the optimal values that maximize the output entropy. For example, a too low laser power would not match the dynamic range of the ADC, whereas a too high power would saturate the PD.

2)	\textsl{Randomness stability}: the generation of unbiased numbers has to be guaranteed for a prolonged and uninterrupted use of the device. Although the operational parameters might stay stable and without drifts, possible errors in the hardware and software implementation and/or failures of the unmonitored components, could introduce bias, correlation or artifacts in the outputted numbers.

We ran the device for 71 days without interruption. To check the operational stability (point (1)), we monitored and recorded the system's operational parameters through the functions presented in Section II.A. To test the randomness stability (point (2)), we applied statistical tests in series and recorded the test results.

The advantage of data logging is that it makes it possible to perform an accurate a-posteriori analysis. In particular, possible failures of the statistical tests can be compared against the values the physical parameters at the time of failure. Concerning this last point, it is well understood that statistical tests on the outputted post-processed data cannot certify the unpredictability of the numbers, as this is meant to be achieved with a theoretical model of the quantum process employed. However, they can detect some deviations from the optimal operating conditions and possible hardware failures. In this sense, the foreground-background monitoring function is useful to enable the comparison of the experimental data against the theoretical model of the generator.

\section{Physical testing}

During the whole testing period, the data of the monitor functions were recorded.
In Table I, we report the mean values and the standard deviations of laser power. This parameter was read by the monitor function from the photodiode embedded into the laser. In Figure 3, we draw its average on intervals of 24 hours.
\begin{table}
\begin{tabularx}{\columnwidth}{p{1 cm} p{3.5 cm} p{2.5 cm} p{2.5 cm}}
&\textbf{Parameter} & \textbf{Value} & \textbf{Unit}\\\hline\hline
&Shannon Entropy & 9.27 $\pm$ 0.02 & bits\\
&Min-Entropy & 8.56 $\pm$ 0.04 & bits\\\hline
\end{tabularx}
\caption{Shannon entropy and min-entropy “values” calculated from the foreground monitor function.}
\label{tab2}
\end{table}

\begin{figure}[t]
\includegraphics[width=\columnwidth]{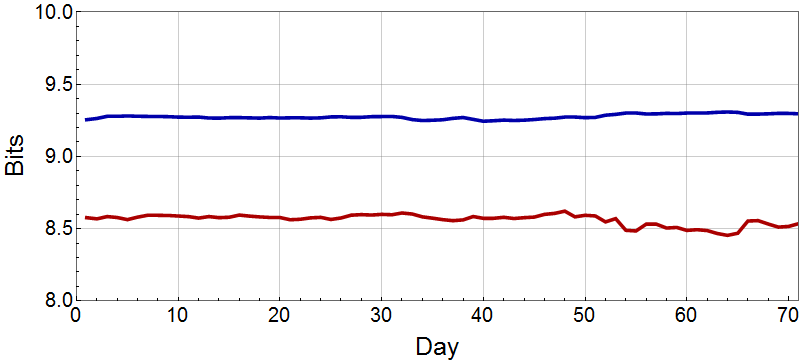}
 \caption{Shannon entropy (blue solid line) and min-entropy (red line) of the QRNG, as evaluated from the foreground data.}
 \label{fig4}
\end{figure}
From this result, it appears that the system is characterized by a stable performance. The laser power features average relative fluctuations of 0.2\%, with a maximal registered deviation of +2.1\% from the mean value.

Such stability was obtained by means of a temperature controller that kept the laser temperature stable during the test. Readouts from the controller monitor function registered a temperature standard deviation of 0.33 °C.

As we already mentioned, it is important that the generator keeps a stable laser power, to obtain a sample distribution that spans the whole ADC range without saturating the PD. 	

The data of the foreground function correspond to the frequency of the digitized interference intensity, $X$, which falls in the interval $[0, 1023]$. The histograms provide the occurrence probabilities $p(X)$ of the events from which the Shannon entropy and the min-entropy are evaluated in real time. These two quantities are, respectively, equivalent to the average and minimal amount of unbiased bits that can be extracted from the raw samples $X$. As mentioned before, the distribution of the $X$ values is not uniform and therefore the entropies are always smaller than the ten bits featured by the ADC.

The benefit of having a temperature-controlled laser with a stable output power is that it creates stable experimental conditions to keep the amount of generated randomness almost constant. This is evident from the data registered for the entropies.

In Table II, we report the entropies mean values and standard deviations for the whole testing period. In Figure 4, the blue and red lines correspond to the daily average of Shannon entropy and min-entropy respectively. We notice that both the entropies feature very limited fluctuations, 0.2\% and 0.45\% for Shannon and min-entropy, respectively, with a maximal registered variation of +0.42\% and +2.43\%, respectively. Hence the QRNG always approaches the highest entropy values allowed by the physics of the process. The high entropy content of the raw samples $X$ made it possible an effective application of the FIR unbiasing filter, as it will be demonstrated by the test results in the next two sections.

\section{Statistical tests}

The task of performing the statistical tests of a random number generator is well studied and a multitude of solutions are available. Well-known batteries commonly applied such as ``FIPS-140-2" \cite{24} or ``Die Hard" \cite{25} are inadequate for our intended analysis, as they are either too weak or too limited in the number of tests. A valid alternative, featuring more stringent tests and an extensive number of batteries, is represented by the Test-U01 and the NIST SP-800-22, which are those we have used to test the QRNG.

\subsection{Test-U01}

The TestU01 suite \cite{26} is acknowledged to be the most stringent collection of tests to assess RNG statistical properties. These tests can be applied singularly or as batteries. For the long term test, we applied the largest battery denominated ``Big Crush". This battery features 106 tests selected to cover the widest spectrum of possible problems. A single run of Big Crush analyzes a dataset of approximately $1.143\times1013$ bits and takes typically 4.5 hours. The ultrafast generation rate of our QRNG enables the generation of this amount of numbers in just 24 minutes. Since this rate exceeds the hard drive writing speed, the datasets were obtained by attaching blocks of numbers with a size of 80 Gb. Each block was acquired in real time, as it was directly written into the volatile memory of the computer, and then moved to a solid state disk to build the dataset. This bottleneck increased the acquisition time to 73 minutes. However, it was still small enough to apply the battery multiple times per day, achieving an unprecedented assurance level for the testing process.

It is worth stressing that the multiple application of a test is fundamental not only to track the generator response, but also to rule out possible ``lucky" or ``unlucky" results that could occur with a single shot application. When a test is applied to a string, one or more p-values are evaluated: if the result is inside the interval $I_{BC} = [0.001, 0.999]$ the test is considered passed. During the testing period the battery was applied 285 times, corresponding to a total of 3.258 Petabit analyzed. The overall number of p-values obtained was 72,390.

To understand how the QRNG performed during the testing period, we studied whether the number $n_{\text{out}}$ of p-values outside $I_{BC}$ was in line with the statistical fluctuations for a sample of that size. Since the probability of failure is $\alpha = 0.002$, we expect $n_{\text{out}} = \alpha \times 72,390 \simeq145$. In addition, we follow \cite{28}, to estimate an acceptance range for $n_{\text{out}}$: by means of the Gaussian approximation to the binomial distribution, we evaluate an acceptance range of  $109 \leq n_{\text{out}} \leq 181$. From the analysis of the test results, we found $n_{\text{out}} = 151$, which is fully compatible with the expectations.

Although the number of suspicious p-values was in line with the expectations, it is necessary to assess the possible presence of a ``catastrophic failure" among the failed tests. For catastrophic failure is intended a p-value very close to 0 or to 1, e.g. $10^{-10}$, whose occurrence cannot be justified in terms of the statistical fluctuations linked to the size of the set of p-values. We verified that no p-value occurred outside the interval $[10^{-6}, 1-10^{-6}]$. For the interval $[10^{-5}, 1-10^{-5}]$, we obtained $n_{\text{out}} = 2$, which is well within the acceptance range $0 \leq n_{\text{out}} \leq 5$.

\begin{figure}[t]
\includegraphics[width=\columnwidth]{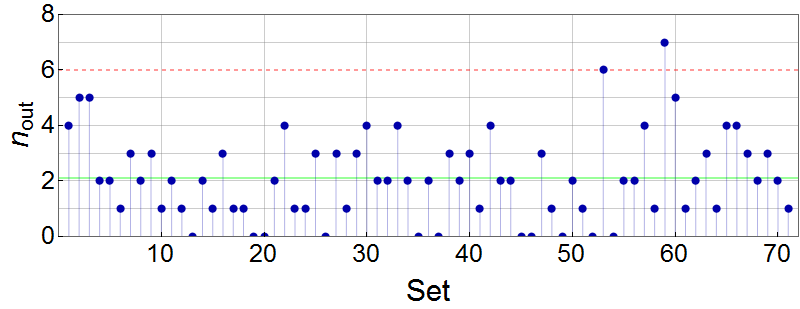}
 \caption{Number of p-values outside the confidence interval in sets containing the results of four consecutive Big Crush battery, which is the typical amount of times the battery was applied per day. The green line represents the daily mean value of $n_\text{out}$. The dashed red line corresponds to the daily acceptance range.}
 \label{fig5}
\end{figure}

The above analysis has assessed that the total number of p-values outside the confidence interval is in line with what is predicted by the theory. However, to check the stationarity of the device, it is relevant to study how these extreme p-values were distributed in time. In particular, we wanted to verify whether clustered failures were present. In fact, a concentration of very high $n_{\text{out}}$ values, in a limited amount of time, would clearly indicate the presence of a physical hardware problem.

In Figure 5, we plot the $n_{\text{out}}$ value after regrouping the p-values in subsets. Each subset contains 1,016 p-values, corresponding to four consecutive runs of the battery. This is indeed the typical number of times the battery was applied per day. The evaluated average of $n_{\text{out}} = 2.11$ (green solid line in the plot) is in line with the expected value of $n_{\text{out}} \simeq 2$ $(0.002 \times 1,016)$. Although we obtained $n_{\text{out}} = 7 $ at subset 59, which is slightly above the threshold of 6 (red dashed line in the plot) we did not consider this fact as suspicious, since the p-values outside the limits belonged each time to different tests. In addition, by cross-checking this result with both the monitor parameters (Figure 3 and 4) and with the results from another test suite (Figure 6), we did not observe any particularly suspicious variation.

It is worth noticing that the values for $n_{\text{out}}$ were evenly distributed during the whole testing period. This indicates that no hardware failures or drifts with a noticeable impact on the output randomness occurred during the whole testing period.

\subsection{NIST SP-800-22}

Although the TEST-U01 represents the state of the art in statistical randomness testing, we analyzed the numbers also with the test suite SP 800-22 developed by the NIST \cite{27,28}. This suite is commonly used for QRNGs and this would ease a comparison between our results and previous works.
\begin{figure}[t]
\includegraphics[width=\columnwidth]{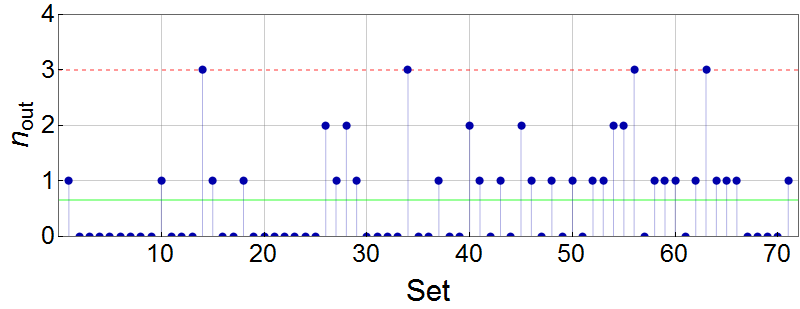}
 \caption{Number of uniformity p-values outside the confidence interval in sets containing the results of forty consecutive applications SP-800-22 suite. The dashed red line corresponds to the daily acceptance range.}
 \label{fig6}
\end{figure}

During the testing period, the battery was applied 2,849 times. For each run we used a 1 Gbit input string and we applied 15 tests. Differently from Big Crush, where each test analyzes different and very long subsamples of the input string, the NIST tests are all applied on the same string. In particular, any new input string is divided into $L = 1,000$ substrings, each 106 bits long, so that each test is applied $L$ times. For the NIST suite the significance level is $\alpha = 0.01$. 
\onecolumngrid
\begin{center}
\begin{figure*}[h]
\includegraphics[width=0.8\textwidth]{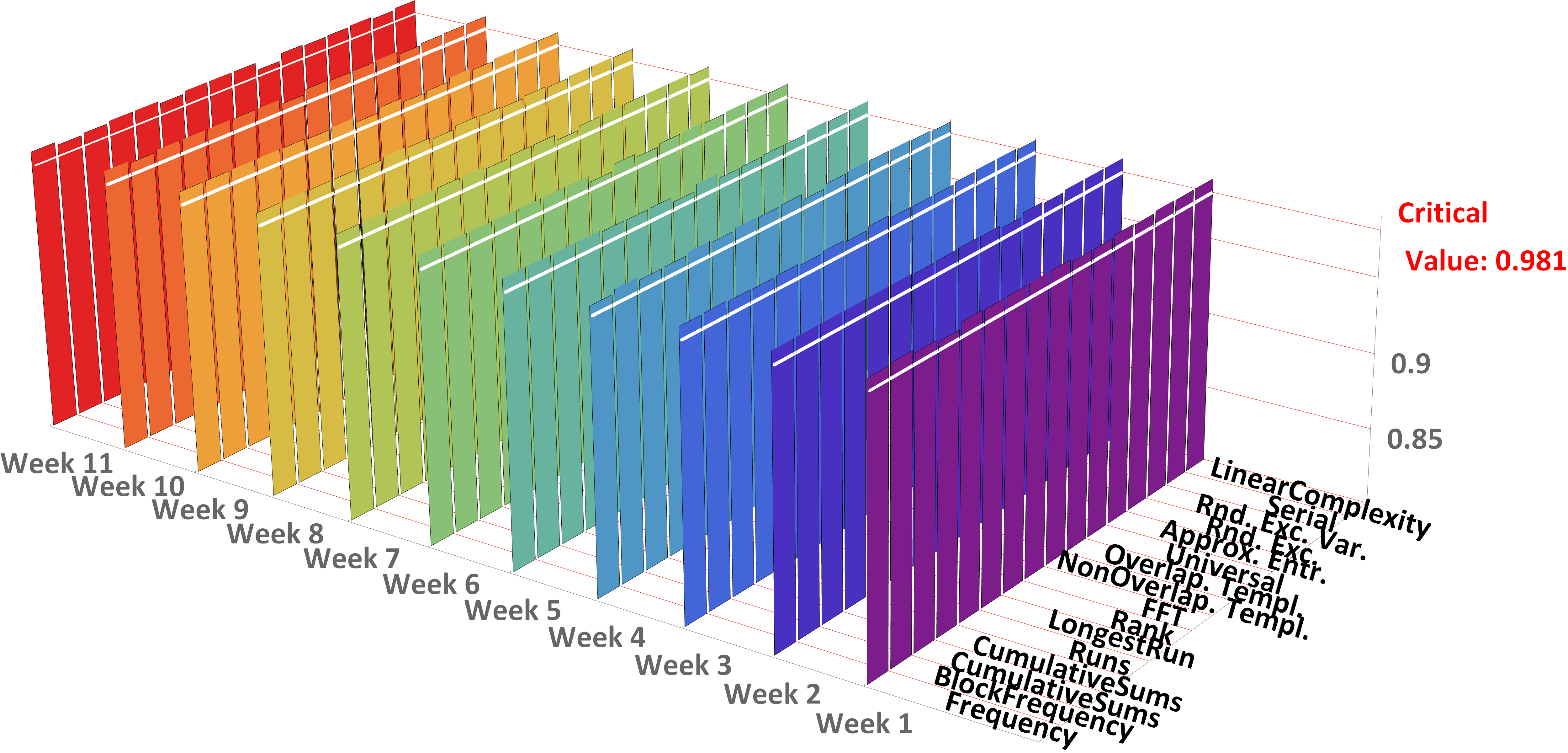}
 \caption{Each column represents the weekly average passing ratio for each test. The passing ratio corresponds to the fraction of 1 Mbit long sub-strings featuring a p-value $ \geq0.01$. For a sample of $L$ = 1,000 bits, the threshold to pass the test is 0.981 and it is represented by the white line on top of each bar plot. For the ``Random Excursion" and ``Random Excursion Variant" tests, $L \simeq 600$ and the threshold is approximately 0.978.}
 \label{fig7}
\end{figure*}
\end{center}
\vspace{-0.5cm}
\twocolumngrid
\noindent
If a test on a given substring yields a p-value $\geq 0.01$, the test is considered passed with confidence 99\%.
If the numbers are statistically sound, the $L$ p-values should follow the uniform distribution.
Hence, for each test, the suite evaluates the sub-string passing ratio and performs a goodness-of-fit (GOF) test on the observed p-value distribution. This second order test is performed by regrouping the $L$ p-values in ten bins of width 0.1, and by computing a $\chi^2$ test statistic. The test on the whole input string is then considered passed if the passing ratio is approximately of 99\%, and if the ``uniformity" p-value associated to the corresponding $\chi^2$ test statistic is $\geq 10^{-4}$.

Each run yield 188 p-values, given that some tests feature many variants. The overall number of the uniformity p-values generated during the 71 days amounts to 535,612. For this sample size, the number $n_{\text{out}}$ of p-values $< 10^{-4}$, is expected to be comprised in the interval $32 \leq n_{\text{out}} \leq 76$ and centred on the value $n_{\text{out}} =54$. We observed $n_{\text{out}} = 46$ which is again fully compatible with the theory.

Also for the NIST suite, we studied the temporal distribution of $n_{\text{out}}$. Each day, the NIST suite was applied 40 times, yielding 7,520 p-values. In Figure 6, the points represent the value of $n_{\text{out}}$ per day. The solid green line corresponds to the daily average value of $n_{\text{out}}$, equal to 0.65, which is close to the expected value $0.75$ $(\simeq 0.0001 \times 7,520)$. The red dashed line corresponds to the threshold level, above which no points were registered. Hence, these results confirmed the conclusions drawn for Big Crush.

As mentioned above, for each test the suite evaluates also the fraction of 1 Mbit long substrings that yielded a p-value larger than 0.01. The suite automatically applies the Gaussian approximation to the binomial distribution. The minimal success fraction to consider the test passed on the whole set of $L$ substrings is 98.1\% (it is worth to specify that for the so called ``Random Excursion" and ``Random Excursion Variant" tests the sample size is $L \simeq 600$, such that the typical threshold fraction is about 97.8\%).

To effectively present the results regarding the passing ratio, we report in Figure 7 the weekly average for each test (since tests ``Non Overlapping Template", ``Random Excursion", ``Random Excursion Variant" and ``Serial" feature different variants, for these tests we report the average on all their variants). As one can appreciate from the figure, the generator kept the averaging passing ratio above the threshold for all tests. Although these results were somehow anticipated from that already presented, the figure is relevant since it confirms that also at the scale of short strings, the generator does not fail to output bits with high statistical qualities.

\section{Conclusion}

We reported the results of an unprecedentedly long 71-day non-stop trial of a QRNG. We showed its capability to sustain uninterrupted operation while providing a continuous, laminar flow of unbiased numbers at 8 Gbit/s.

By analyzing the data recorded from the functions continuously monitoring the physical parameters, we showed that the generator maintained a stationary behavior, without drifting from the optimal experimental conditions. This stability enabled a constant success rate for the most stringent battery of statistical tests, the Big Crush of Test-U01 suite. This is the first time that such an extensive test, with almost 300 applications of the Big Crush battery, is reported. To our knowledge, the only other reported example is with the QRNG developed by PicoQuant to which  the Big Crush test was applied a total of 60 times \cite{29,30}.

Once the generator is provided with a cryptographically-strong randomness extractor, these results show the suitability of the QRNG in cryptographic applications, including quantum key distribution (QKD), for which it is essential to employ random numbers generated from strictly non-deterministic generators. In fact, the ultrafast generation rate makes our QRNG compatible with most of the current QKD systems, which typically feature maximum clock rates around 1 GHz \cite{31}.

We believe that the demonstrated speed, robust design and stable operation of the quantum random number generator will promote it as a valid alternative to the current solutions adopted to generate random numbers.

\begin{acknowledgments}
This project has received funding from the European Union's Horizon 2020 research and innovation programme under the Marie Sklodowska-Curie grant agreement No 750602, project: ``Development of an Ultra-Fast, Integrated, Certified Secure Quantum Random Number Generator for applications in Science and Information Technology" ( UFICS-QRNG ).
\end{acknowledgments}

\end{document}